# Integral surface based Second Order Sliding Mode Controller design for Inverted Pendulum with PD SMC Compensation

Kirtiman Singh, Prabin Kumar Padhy

Stabilization of a nonlinear single stage inverted pendulum is a complicated control problem, as nonlinearity is present inherently and external factors affect the equilibrium position. In this paper, a PD sliding mode controller is connected with Second order PI (Proportional+Integral) sliding mode controller, which is designed to improve the performance for nonlinear state differential equations with unknown parameters. This paper throws light on the sliding surface design and highlights the important features of multiplexing sliding mode control inputs resulting in robustness and higher convergence of output, through extensive mathematical modeling. Simulations and experimental application is done on the system to evaluate the controller for performance, complexity of implementation and also on the impact of the nonlinear IP system on its stability.

**Introduction**: The problem of stabilization of an unstable nonlinear inverted system in its upright position is solved using nonlinear control design technique. Switching systems presents interesting solutions as it is quite simple to implement and provides an optimal solution. They are characterized by changes in the system dynamics, associated by different sets separating each other by a border and the vector fields across the border are directed toward the border itself. Sliding mode control is the special case of variable structure control technique, a switching control strategy. It is a nonlinear control technique which is an invariant control method as it is applicable to both linear and nonlinear systems with unknown parameters and bounded disturbances.

Sliding mode control (SMC) is very suitable for IP system. It provides advantages as it has simple algorithm, invariance & high robustness to parametric uncertainties and external disturbances and provides order reduction. The Inverted Pendulum system is a highly nonlinear dynamic system, challenge to the control technique. Traditional sliding mode control provides an optimum solution to the control of IP system. Sliding mode consist of two phases viz. Sliding surface design and design of a control law to force the system states on the sliding surface. A fatal flaw with traditional SMC is the chattering phenomenon i.e. high frequency oscillations of the controller output resulting in rapid oscillation along the sliding manifold, which is eliminated by using boundary conditions, observer model, or higher order sliding modes. To eliminate the non-robust reaching phase, Integral sliding mode was proposed [15]. In this paper, PI sliding surface based second order sliding mode control is proposed. The proposed method eliminates chattering phenomenon as well as reduces the convergence time. On reducing the finite convergence time, problem of overshoot and singularity arises. Both these problems has been eliminated by the proposed design methodology, where the PD sliding mode controller is used along with it [8, 12]. Control law is derived using reaching laws and equivalent control law [5], which both requires certain knowledge of the system dynamics. In this paper, improved reaching law has been proposed and implemented which markedly improves the response characteristics compared to equivalent control.

This paper has been organized as follows: Section 2 introduces about traditional PD sliding mode control for the IP system model. Section 3 describes the proposed PI sliding surface based sliding mode controller with improved reaching laws. Section 4 discusses the simulation results, with conclusion in Section 5.

**Sliding Mode Control**: Sliding mode control is a special case of a variable structure control. Considering the nonlinear dynamic system
$$\dot{X} = f(x,t,u) + g(x,t).u(t) \quad x \in \mathcal{R}^n, u \in \mathcal{R}^m. \quad (1)$$
subjected to discontinuous feedback
$$u_i = \begin{cases} u_i + (x,t) & if \ s_i(x,t) > 0 \\ u_i - (x,t) & if \ s_i(x,t) < 0 \end{cases} \quad (2)$$
This motion of the system as it slides along the manifold is called sliding mode and the geometrical locus is called sliding surface. Sliding surface is defined as:

$$S(x,t) = (\frac{d}{dt} + \lambda)^{(n-1)}.e \quad (3)$$
A single input nonlinear inverted pendulum's second order equation as:
$$\ddot{\theta} = \frac{mgl\sin(\theta) - m^2l^2\cos(\theta)\sin(\theta)\dot{\theta}^2 + u.ml\cos(\theta)}{m^2l^2\cos^2(\theta) - (I+ml^2)} \quad (4)$$
$$\ddot{\theta} = f(\theta, \dot{\theta}, t) + g(\theta, t).u(t) + d(t) \quad (5)$$
where, $f(\theta, \dot{\theta}, t)$ and $g(\dot{\theta}, t)$ are known nonlinear functions, embodying plant parameters, model uncertainties and unmodelled dynamics.
$d(t)$ = bounded external disturbances.
t = independent time variable.
$\theta$ = Output at any particular time, $\theta(t) \in \mathcal{R}$.
$u(t)$ = control action generated.
The control problem is to asymptotically track the desired response using a discontinuous control law generated by the sliding mode algorithm, in the presence of parameter variations, model uncertainties and bounded external disturbances. To track the desired trajectory, error between the desired trajectory and the reference trajectory is defined as:
$$e(t) = \theta_r(t) - \theta(t) \quad (6)$$

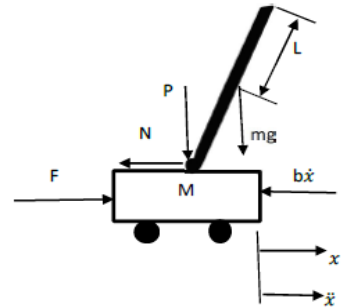

**Fig. 1** Inverted Pendulum Model

**Modified Reaching Law**: Reaching law drives the system to maintain a stable manifold and the sliding phase ensures it to slide to equilibrium. Proposed is the Power rate exponential reaching law for the second order sliding mode as:
$$\ddot{s} = -k_1\dot{s} - k_2s - \varepsilon_1|s|^\alpha sign|s| - \varepsilon_2|s|^\alpha sign|\dot{s}| \quad (7)$$
where, $k_i > 0, \varepsilon_i > 0, 2 < \alpha < 0$.
For the first order PD sliding mode, reaching law used is:
$$\ddot{s} = -k_1 s - \varepsilon_1|s|^\alpha sign(s) \quad (8)$$
In the above equations, proportional term forces the state to approach the switching manifold faster for larger $s$. Power term increases the reaching speed when the state is far away and reduces when the state is near the switching manifold. The power rate sign term generates the control force which guarantees the appearance of a 2-sliding mode attracting the trajectories of the sliding variable dynamics in finite time.

**PI-Second order sliding mode control**: A PI sliding surface is proposed for the second order sliding.
$$s(t) = K_p e(t) + K_i \int_o^t e(\mathcal{T})d\mathcal{T} \quad (9)$$
where $K_p, K_i$ are positive coefficients providing flexibility for sliding surface design and $K_p, K_i \in \mathcal{R}^+$ and $s(t) \in \mathcal{R}$. Assumption being that system is initiated at the region $s(t) > 0$, resulting in increasing $s(t)$ and an insufficient input to drive the error to converge to the sliding manifold. Integral action increases the control action to force the error to the siding manifold. Control action is reduced as $s(t) \to 0$ forcing the system into the sliding mode. Second order sliding surface, denoted by (7), written as:
$$\dot{s}(t) = K_p\dot{e}(t) + K_i e(t) \quad (10)$$
$$\ddot{s}(t) = K_p\ddot{e}(t) + K_i\dot{e}(t) \quad (11)$$
Using equation (4-7) and substituting on (11) results in:
$$\ddot{s}(t) = K_p(\ddot{\theta}_r(t) - \ddot{\theta}(t)) + K_i\dot{e}(t) \quad (12)$$
$$K_p\left(\ddot{\theta}_r(t) - f(\theta,\dot{\theta}) - g(\theta).u(t)\right) + K_i\dot{e}(t) = -k_1\dot{s} - k_2s - \varepsilon_1|s|^\alpha sign(s) - \varepsilon_2|s|^\alpha sign|\dot{s}| \quad (13)$$
Therefore, the control input is given as:
$$u(t) = (K_p g(\theta))^{-1}(-K_p\ddot{\theta}_r(t) + K_p f(\theta,\dot{\theta})) - K_i\dot{e}(t) - k_1\dot{s} - k_2s - \varepsilon_1|s|^\alpha sign(s) - \varepsilon_2|s|^\alpha sign|\dot{s}|) \quad (14)$$

Resulting control input results in an improved response characteristics with fast convergence time. It also results in removal of chattering characteristics. But response due to PI sliding surface leads to a peak overshoot as well as the problem of singularity (Fig 3b). First order derivative of the sliding manifold (PD surface) is represented as

$$\dot{s}(t) = K_p \dot{e}(t) + K_d \ddot{e}(t) \tag{15}$$

$$\dot{s}(t) = K_p\left(\dot{\theta}_r(t) - \dot{\theta}(t)\right) + K_d(\ddot{\theta}_r(t) - \ddot{\theta}(t)) \tag{16}$$

$$K_p\left(\dot{\theta}_r(t) - \dot{\theta}(t)\right) + K_d\left(\ddot{\theta}_r(t) - f(\theta,\dot{\theta}) - g(\theta).u(t)\right) = -k_1 s - \varepsilon_1|s|^\alpha sign(s) \tag{17}$$

$$u(t) = -(K_d g(\theta))^{-1}(-K_d \ddot{\theta}_r(t) + K_d f(\theta,\dot{\theta})) - K_p \dot{e}(t) - k_1 s - \varepsilon_1|s|^\alpha sign(s)) \tag{18}$$

To negate the flaws of the above proposed technique, a PD sliding surface is connected in feedback leading to the removal of the problem of singularity as well as markedly improves the response characteristics (Fig 3c). To eliminate the effect of sign function in control inputs, saturation functions is used as:

$$sat(s) = \begin{cases} 1 & s > \Delta \\ ks & |s| \leq \Delta, k = 1/\Delta \\ -1 & s < -\Delta \end{cases} \tag{19}$$

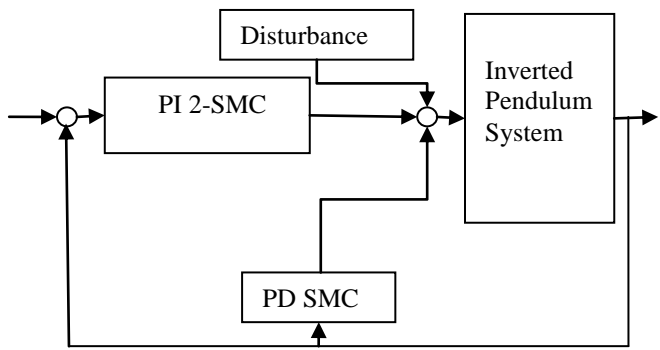

*Fig 2.* Block Diagram for Inverted Pendulum and Control System.

**Simulations**: The output responses for swing up and stabilization of inverted pendulum are compared in the hierarchical manner. Simulations results show the controller is able to achieve desired characteristics successfully. Another point is that higher the order less is the chattering. Also the output has robustness to initial error and parametric uncertainties.

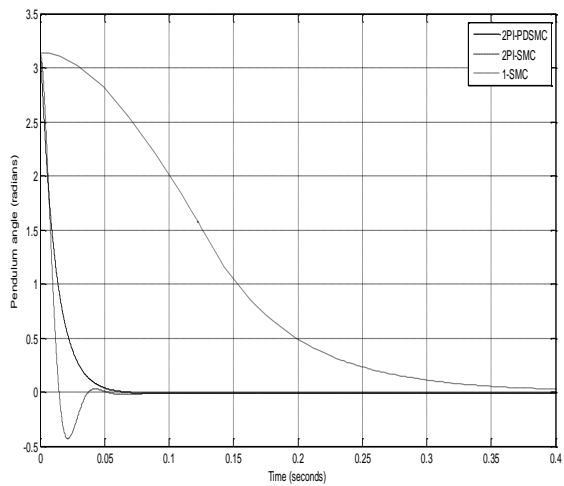

*Fig 3(a).* Output response for PD PI 2-sliding mode control.
  *3(b).* Output response for PI 2-sliding mode control.
  *3(c).* Output response for traditional sliding mode control.

**Conclusion:** PI 2-sliding mode controller along with the PD sliding mode controller is designed for the swing up and stabilization of the Inverted Pendulum with uncertain parameters. Simulations results shows desired output response with reduced chattering and peak overshoot, plus the complete removal of the phenomenon of singularity.